\documentclass[a4paper]{article}
\usepackage[a4paper,margin=3.2cm]{geometry}

\usepackage{calc}
\usepackage{multirow}
\usepackage{times}
\usepackage[T1]{fontenc}
\usepackage{caption}
\usepackage{graphicx}
\expandafter\let\csname equation*\endcsname\relax 
\expandafter\let\csname endequation*\endcsname\relax 
\usepackage{amsmath}
\usepackage{amssymb}
\usepackage{tikz}
\usetikzlibrary{shapes}
\usepackage[hidelinks]{hyperref}
\hypersetup{
 pdfauthor={Nathan Clisby},
 pdftitle={Calculation of the connective constant for self-avoiding walks via the pivot algorithm},
 pdfsubject={},
 pdfkeywords={self-avoiding walk, connective constant, pivot algorithm},
 colorlinks = false,
 pdfborder = {0 0 0}, 
 }

\newtheorem{conjecture}{Conjecture}

\newcommand\tstrut{\rule{0pt}{2.5ex}}
\newcommand\bstrut{\rule[-1.0ex]{0pt}{0pt}}

\newcommand{\tint}{\tau_{\mathrm{int}}}
\newcommand{\cputint}{\widetilde{\tau}_{\mathrm{int}}}

\newcommand{\bt}{\ensuremath{\widetilde{B}}}
\newcommand{\btn}{\ensuremath{\widetilde{B}_N}}

\newcommand{\sci}[1]{\ensuremath{\times 10^{#1}}}

\def\root{\draw[fill] +(0,0) circle (5pt);}
\def\sww{\draw ++(0,0)}

\def\sdw{\draw[style=densely dotted] ++(0,0)}
\def\ew{;}
\def\r{-- ++(1,0)}
\def\l{-- ++(-1,0)}
\def\u{-- ++(0,1)}
\def\d{-- ++(0,-1)}

\begin{document}

\title{Calculation of the connective constant for self-avoiding walks via the pivot algorithm}
\author{Nathan Clisby \\
ARC Centre of Excellence for Mathematics and Statistics of
Complex Systems,\\
Department of Mathematics and Statistics,\\
The University of Melbourne, VIC 3010, Australia}
\date{February 8, 2013}
\maketitle

\begin{abstract}
We calculate the connective constant for self-avoiding walks on the
simple cubic lattice to unprecedented accuracy, using a novel
application of the pivot algorithm. 
We estimate that $\mu = 4.684\, 039\, 931 \pm 0.000\, 000\, 027$.
Our method also provides accurate estimates of the number of
self-avoiding walks, even for walks with millions of steps.
\\ \\
\noindent \textbf{Keywords} self-avoiding walk; connective constant;
Monte Carlo; pivot algorithm; approximate enumeration
\end{abstract}

\section{Introduction\label{sec:intro}}

The self-avoiding walk (SAW) on a regular lattice is an important model in
statistical mechanics with a long history~\cite{Madras1993}. An $N$-step
SAW is a map $\omega$ from the integers $\{0,1,\cdots,N\}$ to sites on
the lattice, with $\omega(0)$ conventionally at the origin, 
$|\omega(i+1) - \omega(i)| = 1$, and $\omega(i) \neq \omega(j) \; \forall i
\neq j$. 
SAW is a topic of much current interest: see
\cite{Bauerschmidt2010} for a recent review of rigorous results, and
\cite{Guttmann2008} for an overview of self-avoiding polygons (SAP) which
has broader scope, including numerical aspects of SAP and to a
lesser extent SAW.

The most important quantities which characterize SAW are the number of
SAW of length $N$, $c_N$, and measures of the size of the walk, such as
the square end-to-end distance. 
The asymptotic behavior of $c_N$ on the simple cubic lattice is believed to be
\begin{align}
c_N &\sim A \mu^N N^{\gamma - 1} \left(1 +
O\left({N^{-\Delta_1}}\right)\right),
\label{eq:cnasympt}
\end{align}
where the connective constant
$\mu$ and amplitude $A$ are lattice dependent, the critical exponent $\gamma$
is universal, and $\Delta_1$ is the exponent of the leading
correction to scaling. There are also sub-leading analytic corrections
to scaling, and a contribution from the so-called anti-ferromagnetic
singularity; see for example~\cite{Clisby2007} for more details on the
asymptotic form of $c_N$.

Enumeration is a particularly powerful method for studying SAW on
two-dimensional lattices, where the finite lattice method is highly
effective~\cite{deNeef1975,deNeef1977,Enting1980}.
The best estimate
for $\mu$ on the square lattice comes from enumerations of self-avoiding
polygons to 130 steps~\cite{Clisby2012}, leading to the highly accurate
estimate $\mu = 2.638\,158\,530\,35(2)$. For the simple cubic lattice,
the best estimate for $\mu$ comes from PERM Monte Carlo
simulations~\cite{Grassberger2005}: $\mu = 4.684\, 038\, 6(11)$. The
most powerful known enumeration method for three-dimensional lattices is
the length-doubling algorithm~\cite{Schram2011}, which combines brute
force enumeration with the inclusion-exclusion principle in a novel way.
SAW on the simple cubic lattice have been enumerated to 36 steps, with
$c_{36} = 2\, 941\, 370\, 856\, 334\, 701\, 726\, 560\, 670$, and $\mu =
4.684\, 040\, 1(50)$~\cite{Schram2011}.

In this paper we will obtain a highly accurate estimate of $\mu$ for SAW
on the simple cubic lattice using a Monte Carlo algorithm. Our method
can also be used to estimate the number of self-avoiding walks.

\section{Method\label{sec:method}}

Our method to calculate $\mu$ for SAW combines four key ideas: 
\begin{enumerate}
\item Use of
the pivot algorithm, the most powerful known method for sampling SAW; 
\item A
novel computer experiment which involves a telescoping sum that
eliminates corrections to scaling;
\item The adoption of scale-free moves to efficiently calculate the
observable of interest;
\item Partitioning CPU time between different sub-problems in an optimal
way.
\end{enumerate}
We now describe each of these aspects in turn.

\subsection{The pivot algorithm\label{sec:pivot}}
The pivot algorithm is an extremely powerful method for sampling SAW in
the canonical ensemble. It was invented by Lal~\cite{Lal1969}, but the
true power of the method was only appreciated after the ground-breaking
work of Madras and Sokal~\cite{Madras1988}. Recently, the
implementation of the pivot algorithm has been improved to make it even
more powerful~\cite{Kennedy2002a,Clisby2010,Clisby2010a}.
The recent improvements make it an extremely attractive prospect to
utilize the pivot algorithm whenever possible.

The pivot algorithm is a Markov chain Monte Carlo algorithm which works
in the set of self-avoiding walks of fixed length, where the elementary
move is a \emph{pivot} as described below. The pivot algorithm generates
a correlated sequence of SAW via the following process:
\begin{enumerate}
\item Select a pivot site of the current SAW according to some
prescription - usually uniformly at random;
\item Randomly choose a lattice symmetry (rotation or reflection);
\item Apply this symmetry to one of the two sub-walks created by splitting
the walk at the pivot site;
\item If the resulting walk is self-avoiding: {\em accept} the pivot
and update the configuration;
\item If the resulting walk is not self-avoiding: {\em reject} the pivot
and keep the old configuration;
\item Repeat.
\end{enumerate}
The pivot algorithm is ergodic, and satisfies the detailed balance
condition which ensures that SAW are sampled uniformly at
random~\cite{Madras1988}.

After a successful pivot, \emph{global} observables, such as 
the square end-to-end distance, change significantly and are
essentially uncorrelated. 
This observation is equivalent to the
statement that the integrated autocorrelation time for a global
observable $A$, $\tint(A)$, is of the same order as the mean time for
a successful pivot. 
In the language of \cite{Madras1988}, once a
successful pivot has been made the resulting configuration is
``essentially new'' with respect to global observables.
For SAW on the simple cubic lattice 
the probability of a pivot attempt being successful is $O(N^{-p})$, 
with $p \approx 0.11$. Therefore
global
observables have $\tint = O(N^p)$; see \cite{Madras1988} for extensive
discussion.

For \emph{local} observables, such as the angle between the 37th and
38th steps of a walk, one may need $O(N)$ successful pivots before the
observable changes. Consequently $\tint = O(N^{1+p})$ for local
observables.

\subsection{Telescoping observable\label{sec:observable}}

\begin{figure}[htb]
\begin{center}
\begin{tikzpicture}[ultra thick,scale=0.70]
\root
\sww \u \l \d \d \d \r \d \l \l \l \u \ew
\begin{scope}[shift={(1,0)}]
\root
\sdw \r \u \u \r \r \d \l \d \r \r \u \ew
\end{scope}
\begin{scope}[shift={(11,0)}]
\root
\sww \u \l \d \d \d \r \d \l \l \l \u \ew
\begin{scope}[shift={(1,0)}]
\root
\sdw \r \u \r \d \d \l \l \l \l \l \u \ew
\end{scope}
\end{scope}
\end{tikzpicture}
\end{center}
\caption{Concatenation of two walks on the square
lattice. On the left the indicator function $B(\omega_1,\omega_2) = 1$, while on the right $B(\omega_1,\omega_2) = 0$.\label{fig:concatenation}}
\end{figure}
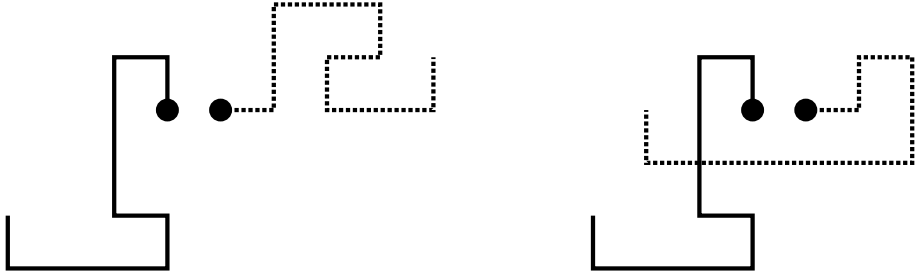

Given walks $\omega_1$ and $\omega_2$, we define a
concatenation operation by placing the root point of $\omega_1$ at the
origin, and the root point of $\omega_2$ at $(1,0,0)$. We denote the
resulting walk as $\omega_1 \circ \omega_2$. Under this definition of
concatenation, walks of $M$ and $N$ steps are fused together to create
a walk of $M+N+1$ steps.
We now define the observable of interest to be the indicator function
defined as follows:
 \begin{align}
 B(\omega_1,\omega_2) &= \begin{cases} 0 & \text{if $\omega_1 \circ
 \omega_2$ not self-avoiding} \\
 1 & \text{if $\omega_1 \circ \omega_2$ self-avoiding} \end{cases} 
 \end{align}
See Fig.~\ref{fig:concatenation} for two examples of concatenation.

The more common definition for concatenation has the root points for
the two walks placed at the origin. 
We use an alternate definition 
because it is straightforward to
calculate the indicator function using our SAW-tree
implementation~\cite{Clisby2010a}.

If we let $\Omega$ be the coordination number of the lattice ($\Omega =
6$ for the simple cubic lattice), we then have
\begin{align}
B_{M,N} &\equiv \text{Mean value of
$B(\omega_1,\omega_2)$ over all pairs of $M$ and $N$ step walks},
\\ &= \langle B(\omega_1,\omega_2) \rangle_{|\omega_1|=M, |\omega_2|=N},
\\ &= \frac{1}{c_M c_N} \sum_{|\omega_1|=M, |\omega_2|=N}
B(\omega_1,\omega_2),
\\ &= \frac{c_{M+N+1}}{\Omega c_M c_N}.
\end{align}
The longest walks which have been exactly enumerated on the simple cubic
lattice have 36 steps~\cite{Schram2011}, and we can recursively exploit
this fact. For convenience we define 
\begin{align}
\btn &\equiv \Omega B_{N,N} = \frac{c_{2N+1}}{c_N^2},
\end{align}
and so
\begin{align}
c_{73} &= \bt_{36} c_{36}^2, \label{eq:btfirst}
\\ c_{147} &= \bt_{73} c_{73}^2 = \bt_{73} \bt_{36}^2 c_{36}^4,
\\ c_{295} &= \bt_{147} \bt_{73}^2 \bt_{36}^4 c_{36}^8,
\\ &\mathrel{\makebox[\widthof{=}]{\vdots}} \nonumber
\\ c_{38797311} &= \bt_{19398655} \bt_{9699327}^2 \cdots
\bt_{36}^{2^{19}}
c_{36}^{2^{20}}.\label{eq:btlast}
\end{align}
Thus, estimates for $\btn$ can be mapped to estimates of the number of walks
$c_N$.
We can then use equation~(\ref{eq:cnasympt}) to estimate $\mu$:
\begin{align}
\mu_N &\equiv c_N^{1/N}
\\ \therefore \log \mu_N &= \frac{1}{N} \log c_N 
\\ &\sim \log \mu + \frac{(\gamma - 1) \log N}{N} + \frac{\log A}{N} +
O\left(N^{-\Delta_1 - 1}\right)
\label{eq:logscaling}
\end{align}
Corrections to scaling vanish with increasing $N$, and estimates for
$\mu_N$ approach $\mu$.

Taking the logarithm of each side of equations
(\ref{eq:btfirst})--(\ref{eq:btlast}), one can see that the contribution
of the $c_{36}$ term remains approximately constant, but the addition of
higher order terms successively eliminate the higher order corrections.
In particular,
\begin{align}
\log \mu_{38797311} &= \frac{1}{38797311}\log\bt_{19398655} +
\frac{2}{38797311}\log\bt_{9699327} + \cdots \nonumber
\\& \hspace{1cm} \cdots + \frac{2^{19}}{38797311}\log\bt_{36}
+ \frac{2^{20}}{38797311}\log c_{36} \label{eq:mu38797311}
\end{align}

The approach described here may be thought of as a
``divide-and-conquer'' algorithm, where a long SAW is successively split
into halves. This is in stark contrast to typical growth algorithms such
as PERM, where SAW (and other combinatorial objects) are incrementally
built up step by step.

\subsection{Scale-free moves\label{sec:scalefree}}

In order to accurately estimate $\mu$ from
equation~(\ref{eq:mu38797311}), we must find an efficient way 
to estimate $\btn$. We estimate $\btn$ by sampling pairs of SAW of
length $N$ via the pivot algorithm, and then $\btn$ is the time
average of $\Omega B(\omega_1,\omega_2)$.
The observable $B$ is \emph{not} a global observable in the same sense as,
for example,
the square end-to-end distance: it clearly depends strongly on the
details of the structure of each walk close to the concatenation joint.

We now present a simple yet subtle argument to show that if we naively
sample pivot sites uniformly at random, then $\tint$ for $B$ will be
$O(N)$. We will assume throughout that we are considering pairs of walks
of length $N$.

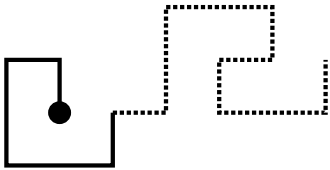
\begin{figure}[htb]
\begin{center}
\begin{tikzpicture}[ultra thick,scale=0.70]
\root
\sww \u \l \d \d \r \r \u \ew
\begin{scope}[shift={(1,0)}]
\sdw \r \u \u \r \r \d \l \d \r \r \u \ew
\end{scope}
\end{tikzpicture}
\end{center}
\caption{Minimal trapped walk of seven steps on the square lattice (solid line) with a possible
extension (dashed line). \label{fig:trapped}}
\end{figure}

First, let us define zero atmosphere SAW as those self-avoiding walks
for which one of the ends has all neighboring sites occupied. It is
well known that zero atmosphere walks have positive density in the set
of all walks (see e.g. \cite{Owczarek2008}). 
We denote a zero atmosphere SAW as ``minimal'' if, starting from
the end, we visit all of the neighbors of the end in the fewest possible
number of steps.  Minimal zero atmosphere walks also have positive
density in the set of SAW.  E.g. for the square lattice, the density of
minimal zero atmosphere SAW which start with the seven steps in
Fig.~\ref{fig:trapped} is bounded below as the SAW length $N \rightarrow
\infty$.

Our ensemble is pairs of SAW, each of fixed length. Suppose we were
to sample pivot sites uniformly at random, so generating a Markov chain.
Assume we have equilibrated the Markov chain so that we are guaranteed
to be sampling from the equilibrium distribution. If we were to choose
a random time in the Markov chain, the probability of choosing a minimal
zero atmosphere walk is then $O(1)$. However, the probability
that the next pivot site chosen could change the value of the atmosphere
is $O(1/N)$. Therefore, in this case $B$ will, on average, remain
zero for $O(N)$ time steps in the Markov chain. For the observable $B$, the
contribution of zero atmosphere walks ensures that it must take time
$O(N)$ to achieve an essentially new configuration. Thus, $\tint(B) = O(N)$
when pivots are sampled uniformly at random. Note that this effect is
actually quite subtle, as although zero atmosphere walks have positive
density, in practice this density is small. Thus the contribution of
these configurations to $\tint(B)$ is small in practice until $N$ is of the
order of thousands or tens of thousands.

However, it is possible to dramatically improve the integrated
autocorrelation time for $B$, and hence the accuracy of our estimate of
$\btn$. The key
point is that the concatenation operation introduces a new,
important length scale into the system, namely the distance from the
concatenation joint to internal sites of the walk.
$B$ depends strongly on the structure of the walk according to this
distance. 
We make the following conjecture.
\begin{conjecture}\label{conjecture}
Suppose we have an observable for a polymer system that depends on a
single internal
distance, $L$. Then the integrated autocorrelation time for this
observable is of the same order as the time it takes to make successful
pivots at all length scales with respect to this
distance.
\end{conjecture}
To be concrete, if $L$ is the distance
from an internal site to the concatenation joint, then we believe that
an essentially new configuration with respect to $B$ is obtained once
pivots have been made at length scales $L$ of order $1,2,4,8,16,\cdots,N$.

By choosing pivot sites uniformly at random with respect to $\log L$, we
therefore expect that there is only at most a $\log N$ penalty for the
integrated autocorrelation time for $\btn$ as compared to a global
observable, i.e. $\tint(B) = O(N^p \log N)$. N.B., since the CPU time
per attempted pivot for the SAW-tree implementation is $O(\log
N)$~\cite{Clisby2010a}, this means that in CPU units $\cputint(B) =
O(N^p \log^2 N)$. 

\subsection{Experimental design\label{sec:design}}

To estimate $\mu$ we must calculate each of the terms in equation
(\ref{eq:mu38797311}). We do so by running separate Monte Carlo
simulations for pairs of walks of length $36$, $73$, $\cdots$,
$19398655$, in order to calculate $\btn$. 
Since it takes CPU time $O(\log N)$ to make a pivot attempt, and CPU
time $O(\log N)$ to calculate 
$B$, we choose to sample $B$ for every time step in the Markov
chain. The procedure we used was: 
\begin{enumerate}
\item Use the pseudo\_dimerize procedure of \cite{Clisby2010a} to generate
two initial $N$-step SAW configurations.
\item Initialize Markov chain by performing at least $20 N$ successful
pivots
on each SAW. Pivot sites are sampled uniformly at random. The stopping
criterion must be based on the number of attempted pivots so as not to
introduce bias.
\end{enumerate}
Our sampling procedure for $B$ is then:
\begin{enumerate}
\item Select one of the two walks uniformly at random.
\item Select a pivot site on this walk by generating a pseudorandom
number $x$ between 0 and $\log N$, and let pivot site $j = \lfloor e^x
\rfloor$.
\item Attempt pivot move, update walk if result is self-avoiding.
\item Randomly pivot each of the walks around their root points. These
pivots are always successful.
\item Calculate $B(\omega_1,\omega_2)$, and update our estimate of $\btn$.
\item Repeat.
\end{enumerate}

Our goal is to optimally partition CPU time amongst the terms in
equation (\ref{eq:mu38797311}), in order to minimize the overall error
in our estimate of $\mu$.
The terms in equation (\ref{eq:mu38797311}) approach
 $\frac{2}{N} \log \btn$ for large $N$. We have
\begin{align}
\btn &= \frac{c_{2N+1}}{c_N^2} \sim \frac{A \mu^{2N+1}
(2N+1)^{\gamma-1}}{A^2 \mu^{2N} N^{2(\gamma -1)}} \sim C N^{1-\gamma},
\\ \therefore \frac{1}{N} \log\btn &\sim \frac{1-\gamma}{N} \log N +
O(1/N).
\end{align}
The $1/N$ factor on the right hand side of the above equation dominates
the increase in integrated autocorrelation time in CPU units for
$B$. Therefore if we were to invest the same CPU time in each term
of equation (\ref{eq:mu38797311}), the contributions to the error would
diminish with increasing $N$!

\begin{figure}[htb]
\begin{center}
\includegraphics[scale=1.0]{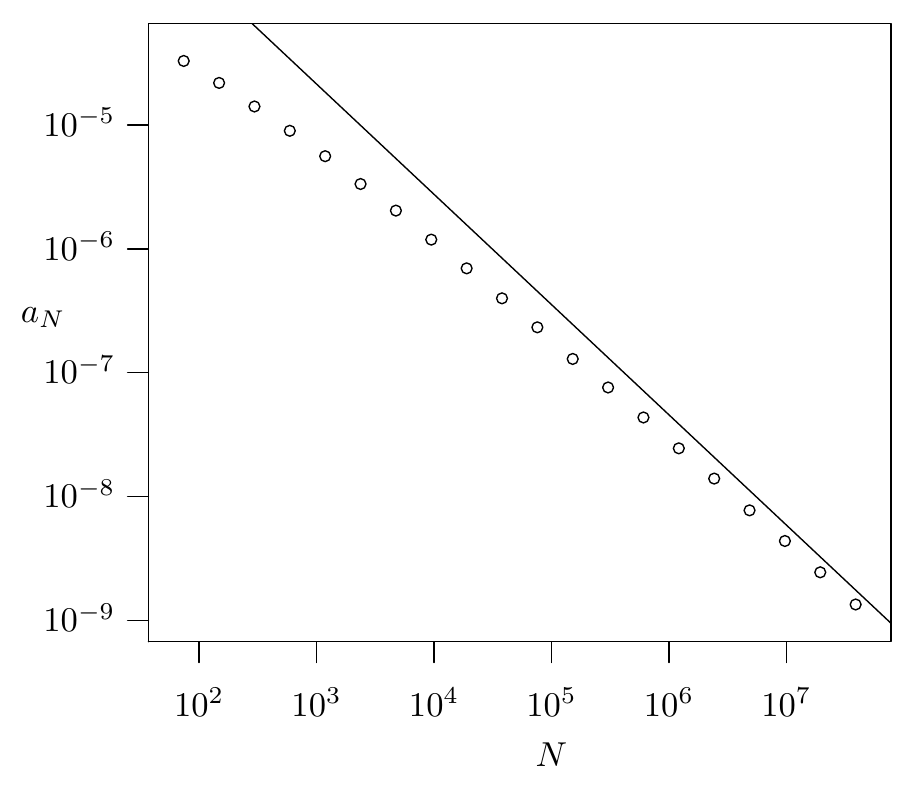}
\vspace{-2ex}
\end{center}
\caption{Measured values of $a_N$, which measures the expected error of
contributions to equation~(\ref{eq:mu38797311}), in units of
$\sqrt{\text{seconds}}$. A line of slope $(-1+(p+\gamma-1)/2) \approx
-0.87$ is included
in the plot for comparison.\label{fig:aN}}
\end{figure}

To minimize overall
statistical error we now perform a short test run of CPU time $t_0$ for
each length, determining the constants $a_N$ in
\begin{align}
\sigma\left(\frac{1}{N} \log \btn\right) &= \frac{a_N}{\sqrt{{t_0}}}.
\end{align}
We show these measured values of $a_N$ in Fig.~\ref{fig:aN}.
However, we can also express $\sigma$ in terms of the variance of
$B$ and the integrated autocorrelation time of the algorithm.
Assuming that Conjecture~\ref{conjecture} is correct, modulo logarithmic
factors we obtain the following expression for $a_N$:
\begin{align}
a_N &\sim N^{-1 + (p+\gamma - 1)/2} \approx N^{-0.87}.
\end{align}
In Fig.~\ref{fig:aN}, it is clear that $a_N$ decays as a power law with $N$,
as expected. 
By inspection, $a_N$ follows the predicted power law behavior quite
closely, and thus Fig.~\ref{fig:aN} provides strong
numerical support for Conjecture~\ref{conjecture}.

We
then fix the total running time for our computer experiment at $t$. The
(statistical) square error in our estimate for $\mu$ is then
\begin{align}
\sigma^2 &= \sum \frac{a_i^2}{t_i}, \hspace{2ex} \text{subject to $\,
t=\sum t_i$}.
\end{align}
The optimal choice of $t_i$ to minimize $\sigma^2$ is then
\begin{align}
t_i &= \frac{a_i}{\sum a_i} t,
\end{align}
and the optimal value for the error is
\begin{align}
\sigma = \frac{\sum a_i}{\sqrt{t}}.
\end{align}
In practice, we did not rigorously apply this prescription to the
longest walks, and instead spent at minimum 1\% of the CPU time at each
length. 

Almost all of the computational effort is spent on the $\bt_{36}$ and
$\bt_{73}$ terms in equation (\ref{eq:mu38797311}). The higher order
terms reduce the corrections to scaling, and essentially eliminate the
systematic error in our estimate of $\mu$.

\section{Results and Analysis}

The analysis for this computer experiment is remarkably simple. It is an
extremely rare example of a problem in lattice statistical mechanics for
which we have strong evidence that systematic errors are negligible.
Hence the confidence intervals we report are purely statistical. 

We ran the computer experiment for a total of $60\, 000$ CPU hours on
SunFire X4600M2 machines with 2.3GHz AMD Opteron CPUs. 

In Table~\ref{tab:cn} we report our estimates for $\btn$, and thence
our estimates for $c_N$ from
equations~(\ref{eq:btfirst})--(\ref{eq:btlast}).
Note that the estimates for $c_N$ are highly correlated.
The error in the mantissa is given in the final column; for example,
from Table~\ref{tab:cn} we estimate
that $c_{38797311} = 6.6 \times 10^{26018276}$, with the confidence
interval of the mantissa being $(5.3,8.2)$. 
This is a direct estimate
from our $\btn$ values: it is \emph{not} an extrapolation, and 
the reported error is purely statistical.
As a technical aside, the error estimates for $\btn$ in
Table~\ref{tab:cn} are approximately
constant for $N \geq 151551$ because we invested the same 
percentage of CPU time in
each of these cases.

\begin{table}[htb]
\begin{center}
\begin{tabular}{rllc}
\hline
\tstrut \bstrut {$N$} & \multicolumn{1}{c}{$\bt_{(N-1)/2}$} &
\multicolumn{1}{c}{$c_N$} \hspace{5ex} & \multicolumn{1}{c}{
$c_N$ mantissa interval } \\
\hline
 \tstrut 73 & 2.47267030(65)& 2.139271\sci{49} & (2.139270, 2.139271) \\
 147 & 2.20753977(91)& 1.010276\sci{99} & (1.010275, 1.010277) \\
 295 & 1.9740142(14) & 2.014793\sci{198} & (2.014790, 2.014796) \\
 591 & 1.7668271(18) & 7.172241\sci{396} & (7.172218, 7.172264) \\
 1183 & 1.5823991(25) & 8.140025\sci{793} & (8.139971, 8.140078) \\
 2367 & 1.4178577(36) & 9.394724\sci{1587} & (9.394599, 9.394850) \\
 4735 & 1.2708081(58) & 1.121626\sci{3176} & (1.121595, 1.121656) \\
 9471 & 1.1392521(81) & 1.433230\sci{6352} & (1.433151, 1.433308) \\
 18943 & 1.0214669(91) & 2.098243\sci{12704} & (2.098013, 2.098474) \\
 37887 & 0.9159517(92) & 4.032592\sci{25408} & (4.031706, 4.033477) \\
 75775 & 0.8214372(97) & 1.335804\sci{50817} & (1.335217, 1.336391) \\
 151551 & 0.736643(10) & 1.314444\sci{101634} & (1.313290, 1.315600) \\
 303103 & 0.660651(10) & 1.141449\sci{203268} & (1.139445, 1.143457) \\
 606207 & 0.592531(11) & 7.720126\sci{406535} & (7.693038, 7.747310) \\
 1212415 & 0.531449(11) & 3.167451\sci{813071} & (3.145262, 3.189797) \\
 2424831 & 0.476654(11) & 4.782146\sci{1626142} & (4.715379, 4.849858) \\
 4849663 & 0.427497(11) & 9.776394\sci{3252284} & (9.505309, 1.005521) \\
 9699327 & 0.383408(12) & 3.664531\sci{6504569} & (3.464124, 3.876531) \\
19398655 & 0.343919(12) & 4.618409\sci{13009138} & (4.127077, 5.168235) \\
38797311 & 0.308455(11) & 6.579250\sci{26018276} & (5.253839, 8.239029) \\
\hline
\end{tabular}
\end{center}
\caption{Estimates of $\btn$ and $c_N$ with statistical errors.\label{tab:cn}}
\end{table}

In our analysis for $\mu$ we utilize an estimate for the critical exponent $\gamma$ from a Monte
Carlo computer experiment~\cite{Clisby2013}: $\gamma = 1.15696(1)$. In
addition, we utilize the estimate of the critical amplitude $A =
1.215(2)$ from \cite{Clisby2007}. We do this by setting $\gamma^* =
1.15696$, $A^* = 1.215$, and forming the improved
estimates
\begin{align}
\log \mu^*_N &= \log\mu_N - \frac{(\gamma^*-1)\log N}{N} - \frac{\log
A^*}{N}.
\end{align}
We denote the errors in the utilized estimates as $\sigma_\gamma =
0.00001$ and
$\sigma_A = 0.002$.
In the limit of large $N$, $\mu^*_N$ will then have
the following contributions to the systematic error:
\begin{align}
&\frac{\mu \sigma_\gamma \log N}{N}, \frac{\mu \sigma_A }{A N},
O(N^{-\Delta_1-1}).
\end{align}
The $\Delta_1$ term comes from the leading order correction in
equation~(\ref{eq:logscaling}). From \cite{Clisby2010} we have $\Delta_1 = 0.528(12)$. The constant of this term is indeterminate, but we will 
see that it cannot be so large so as to interfere with our
estimates.

Our estimates for $\mu$ are
collected in Table~\ref{tab:data}. The $\mu^*_N$ estimates 
rapidly converge with increasing $N$, which indicates that for the
largest values of $N$ systematic errors are negligible. We can also see
from the table that the statistical error, $\sigma(\mu^*_N)$, is
dominated by the low order terms. Finally, it is clear that the
contributions from the errors of the $\gamma^*$ and $A^*$ terms are much
smaller than the statistical error for large $N$.

One additional point is that for the largest values of $N$,
$N^{-\Delta_1-1}$ is of the order of $10^{-11}$. In principle, this term
could have a large constant and result in a large and unknown systematic
error. In practice, because of the smooth convergence of our estimates
we know that the constant cannot be large, and hence contributions from
this term to $\mu^*_N$ are negligible for large $N$.

\begin{table}[htb]
\begin{center}
\begin{tabular}{rlllll}
\hline
\tstrut \bstrut {$N$} & \multicolumn{1}{c}{$\mu^*_N$} &
\multicolumn{1}{c}{$\sigma(\mu^*_N)$} & \multicolumn{1}{c}{${\mu
\sigma_\gamma \log N}/{N}$} & \multicolumn{1}{c}{${\mu \sigma_A }/{(A
N)}$} & \multicolumn{1}{c}{$N^{-\Delta_1-1}$} \\
\hline
\tstrut 73 & 4.68373253707 & 1.70\sci{-8} & 2.79\sci{-6} & 1.07\sci{-4} & 1.60\sci{-3} \\
147 & 4.68392658487 & 2.13\sci{-8} & 1.60\sci{-6} & 5.28\sci{-5} & 5.61\sci{-4} \\
295 & 4.68400034315 & 2.40\sci{-8} & 9.06\sci{-7} & 2.62\sci{-5} & 1.97\sci{-4} \\
591 & 4.68402683289 & 2.53\sci{-8} & 5.07\sci{-7} & 1.31\sci{-5} & 6.96\sci{-5} \\
1183 & 4.68403589477 & 2.60\sci{-8} & 2.80\sci{-7} & 6.52\sci{-6} & 2.46\sci{-5} \\
2367 & 4.68403883775 & 2.64\sci{-8} & 1.54\sci{-7} & 3.26\sci{-6} & 8.68\sci{-6} \\
4735 & 4.68403971655 & 2.68\sci{-8} & 8.37\sci{-8} & 1.63\sci{-6} & 3.07\sci{-6} \\
9471 & 4.68403994072 & 2.70\sci{-8} & 4.53\sci{-8} & 8.14\sci{-7} & 1.08\sci{-6} \\
18943 & 4.68403997588 & 2.71\sci{-8} & 2.44\sci{-8} & 4.07\sci{-7} & 3.84\sci{-7} \\
37887 & 4.68403996593 & 2.71\sci{-8} & 1.30\sci{-8} & 2.04\sci{-7} & 1.36\sci{-7} \\
75775 & 4.68403995443 & 2.71\sci{-8} & 6.95\sci{-9} & 1.02\sci{-7} & 4.79\sci{-8} \\
151551 & 4.68403994395 & 2.71\sci{-8} & 3.69\sci{-9} & 5.09\sci{-8} & 1.69\sci{-8} \\
303103 & 4.68403993749 & 2.72\sci{-8} & 1.95\sci{-9} & 2.54\sci{-8} & 5.99\sci{-9} \\
606207 & 4.68403993406 & 2.72\sci{-8} & 1.03\sci{-9} & 1.27\sci{-8} & 2.12\sci{-9} \\
1212415 & 4.68403993235 & 2.72\sci{-8} & 5.41\sci{-10}& 6.36\sci{-9} & 7.49\sci{-10}\\
2424831 & 4.68403993145 & 2.72\sci{-8} & 2.84\sci{-10}& 3.18\sci{-9} & 2.65\sci{-10}\\
4849663 & 4.68403993096 & 2.72\sci{-8} & 1.49\sci{-10}& 1.59\sci{-9} & 9.36\sci{-11}\\
9699327 & 4.68403993069 & 2.72\sci{-8} & 7.77\sci{-11}& 7.95\sci{-10}& 3.31\sci{-11}\\
19398655 & 4.68403993058 & 2.72\sci{-8} & 4.05\sci{-11}& 3.97\sci{-10}& 1.17\sci{-11}\\
38797311 & 4.68403993052 & 2.72\sci{-8} & 2.11\sci{-11}& 1.99\sci{-10}& 4.14\sci{-12}\\
\hline
\end{tabular}
\end{center}
\caption{Estimates of $\mu^*_N$ with statistical error
$\sigma(\mu^*_N)$, and contributions to the systematic error.\label{tab:data}}
\end{table}

We thus conclude that the estimate $\mu^*_{38797311}$ has negligible
systematic error, and hence adopt this as our best estimate for $\mu$.
Our final estimate is $\mu = 4.684\, 039\, 931(27)$. 

Note, we \emph{could} have avoided the use of previous estimates of
$\gamma$ and $A$, had the calculation of $\btn$ been extended to larger
$N$. This was not done because for $N$ of the order of 100 million or
so, both memory management and initialization time become significant
but not insurmountable issues for the simulation of SAW using the
SAW-tree implementation~\cite{Clisby2010a}.

\section{Discussion}

As noted in the introduction, for the calculation of $\mu$ the approach
which is most competitive with the algorithm presented in this paper is
PERM~\cite{Grassberger2005}, where the estimate $\mu = 4.684\,
038\, 6(11)$ was obtained. Our error bar is approximately 40 times
smaller, which is clearly a significant improvement upon the previous
state of the art.
Other approaches to the calculation of $\mu$ worth noting are the method of
atmospheres~\cite{Rechnitzer2002}, and the Berretti-Sokal
algorithm~\cite{Berretti1985}.

We note in passing that the method of atmospheres could be combined with
the pivot algorithm and scale-free moves to obtain an accurate estimate
for $\mu$. We will not go into any depth, but the method of atmospheres
corresponds to estimating
\begin{align}
\frac{c_{N+K}}{c_N c_K} &\sim \frac{A \mu^{N+K}}{A \mu^N c_K} = \frac{\mu^K}{c_K},
\end{align}
for small, fixed $K$, and in the limit $N \rightarrow \infty$.  From this
expression one can then estimate $\mu$ once corrections-to-scaling have
been
taken into account.  Despite being more accurate than previous methods,
it is an order of magnitude less accurate than the method described
here. This is because 
the mean
CPU time per pivot attempt is $O(\log N)$
for the SAW-tree implementation.
For the atmospheric sampling method, the dominant error comes from
sampling walks in the large $N$ limit, while for the method described in
this paper the dominant error term comes from sampling short walks (in
our case, with $N = 36$).

On the topic of approximation enumeration of SAW beyond the limit of exact
enumeration, there have been a number of papers in recent years.
Approaches include incomplete enumeration~\cite{Sumedha2005}, flatPERM
and flatGARM~\cite{JansevanRensburg2010a}, stochastic
enumeration~\cite{Rubinstein2012}, and the multicanonical Monte Carlo
method~\cite{Shirai2012}.
The relative
advantage of our approach
is significant for small $N$, e.g Shirai and Kikuchi~\cite{Shirai2012}
obtained $c_{256} =
6.2(4) \times 10^{108}$ for the square lattice, while for comparison we
found $c_{295} = 2.014793(3) \times 10^{198}$ on the simple cubic
lattice. For larger $N$, the relative advantage of our method
increases, since to generate a SAW using an incremental growth method
takes CPU time at least $O(N)$. This factor of $N$ becomes prohibitively
large when $N$ is of the order of millions.

It is not clear to us if our approach could be adapted to other
approximate enumeration problems, or to estimations of the free energy
for other models in statistical mechanics. The general principles of
``divide-and-conquer'' and the use of global moves in the canonical
ensemble may be of wider use, or it may be that SAW is a particularly
favorable model.

We consider Fig.~\ref{fig:aN} to be strong evidence in 
favor of the correctness of Conjecture~\ref{conjecture}.
We therefore expect that the use of scale-free moves for the simulation of
polymers will prove useful
in other contexts where there are additional length scales. For example,
in the cases of star polymers or confined polymers. We will explore
this idea further in a future paper where we will also derive an
estimate of the critical exponent $\gamma$~\cite{Clisby2013}.

In future, our implementation of the SAW-tree could be optimized for the
non-uniform selection of pivot sites according to our scale-free
prescription. In particular, there is no reason a pivot being performed
near the end of a walk should take mean CPU time $O(\log N)$. It is
possible to arrange the binary tree data structure so that this
operation would take time $O(1)$. One natural way of doing this would be
to use a splay tree~\cite{Sleator1985}, which would dynamically adjust
to form an optimal tree structure for any choice of pivot site sampling
distribution.

We could also obtain a constant factor improvement, if it were
possible to efficiently forbid configurations with immediate returns at
the concatenation joint.

Finally, it is certainly possible to apply this approach to other
lattices. Unfortunately, in the case of the square lattice the finite
lattice method enumerations of polygons provide estimates for
$\mu$~\cite{Clisby2012}
which are approximately 2 orders of magnitude more accurate than our
method. However, for three-dimensional lattices such as the body
centered cubic lattice and the face centered cubic lattice, our method
will allow for much more accurate calculations of $\mu$ than are
currently available.

\section{Conclusion}

We have applied the pivot algorithm to calculate the connective constant
for self-avoiding walks on the simple cubic lattice, obtaining $\mu =
4.684\, 039\, 931(27)$. Our approach may also be used to derive
extremely accurate estimates for the number of self-avoiding walks.
The power of our approach derives from the
application of an efficient global move (the pivot algorithm), use of an
observable which is calculated through a divide-and-conquer approach,
and from the application of scale-free moves. We hope that these key
ideas may prove useful in other contexts.

\section*{Acknowledgments}
Financial support from the ARC Centre of Excellence for Mathematics and Statistics of
Complex Systems is gratefully acknowledged.

\end{document}